\documentclass[aps, prd, superscriptaddress, nofootinbib, preprintnumbers]{revtex4}
\usepackage{amsmath}
\usepackage{graphicx}

\newcommand{\lesssim}{\mathrel{\mathpalette\vereq<}}
\newcommand{\gtrsim}{\mathrel{\mathpalette\vereq>}}

\newcommand{\chushi}[1]{}

\begin{document}
 \preprint{MISC-2011-15}
 \title{{\bf Techni-dilaton signatures at LHC}
 \vspace{5mm}}
\author{Shinya Matsuzaki}\thanks{
      {\tt synya@cc.kyoto-su.ac.jp} }
      \affiliation{ Maskawa Institute for Science and Culture, Kyoto Sangyo University, Motoyama, Kamigamo, Kita-Ku, Kyoto 603-8555, Japan.}
\author{{Koichi Yamawaki}} \thanks{
      {\tt yamawaki@kmi.nagoya-u.ac.jp}}
      \affiliation{ Kobayashi-Maskawa Institute for the Origin of Particles and 
the Universe (KMI) \\ 
 Nagoya University, Nagoya 464-8602, Japan.}
\date{\today}

\begin{abstract}
We explore discovery signatures of techni-dilaton (TD) at LHC. 
The TD was predicted long ago as a composite pseudo Nambu-Goldstone boson (pNGB)   
associated with the spontaneous breaking of the approximate 
scale symmetry in the walking technicolor (WTC) 
(initially dubbed  ``scale-invariant technicolor''). 
Being pNGB, whose mass arises from the explicit scale-symmetry breaking due to the spontaneous breaking itself (dynamical mass generation), 
the TD as a composite scalar should have a mass $M_{\rm TD}$  lighter than other techni-hadrons,
say $M_{\rm TD} \simeq 600$ GeV for the
typical WTC model, which is well in the discovery range of the ongoing LHC experiment. 
We develop a spurion method of nonlinear realization to calculate the TD couplings to the standard model (SM) particles and  
explicitly evaluate the TD LHC production cross sections at $\sqrt{s}=7$ TeV 
times the branching ratios  in terms of $M_{\rm TD}$ as an input parameter for the region  
 200 GeV $< M_{\rm TD} <$ 1000 GeV in the typical WTC models. 
 It turns out that the TD signatures 
 are quite different from those of the SM Higgs: 
In the one-doublet model (1DM) all the cross sections including the $WW/ZZ$ mode are 
suppressed compared to those of the SM Higgs due to the suppressed TD couplings, 
while in the one-family model (1FM) 
all those cross sections get highly enhanced 
 because of 
the presence of extra colored fermion (techni-quark) contributions.  
We compare the ${\rm TD} \to WW/ZZ$ signature with the recent ATLAS and CMS bounds and 
find  that in the case of 1DM the signature is consistent over the whole mass range 200 GeV $< M_{\rm TD} <$ 1000 GeV due to the large suppression of 
TD couplings, and by the same token  
the signal is too tiny for the TD to be visible through this channel at LHC.  
As for the 1FMs, on the other hand,  a severe constraint is given on the TD mass to exclude the TD with mass 
$\lesssim 600$ GeV, 
which, however, would imply an emergence of somewhat 
dramatic excess as the TD signature at $600 \, {\rm GeV} \lesssim M_{\rm TD} <$ 1000 GeV in the near future.   
We further find a characteristic signature coming from the $\gamma\gamma$ mode in the 1FM. 
In sharp contrast to the SM Higgs case, 
it provides % the 
highly enhanced cross section $\sim 0.10$--$1.0$ fb at around the TD mass $\simeq 600$ GeV, 
which is large enough to be discovered during the first few year's run at LHC.

\end{abstract}
\maketitle

\section{Introduction} 

Large Hadron Collider (LHC) has started setting the strong constraints on the standard model (SM) Higgs boson, the key particle 
responsible for  the origin of mass in the context of the SM. 
The recent data from the ATLAS~\cite{ATLAS} and CMS~\cite{CMS} experiments suggest that 
the SM Higgs boson is unlikely for the mass range  
as low as the electroweak (EW) scale, which may suggest that 
there might exist  certain composite dynamics for the origin of mass due to the  
strongly coupled theories like technicolor (TC)~\cite{Farhi:1980xs}. 
Actually, in contrast to the original TC~\cite{Weinberg:1975gm}, a Higgs-like object, techni-dilaton (TD),  was predicted as a composite scalar, a  pseudo Nambu-Goldstone boson (pNGB)  
associated with the spontaneously broken approximate scale symmetry 
 in the walking TC (WTC), initially dubbed ``scale-invariant TC''~\cite{Yamawaki:1985zg,Bando:1986bg}. 
Thus clarifying the TD signature at LHC is the urgent task in 
searching for Higgs-like particle  at the ongoing LHC, which is the target of this paper.

The original version of TC~\cite{Weinberg:1975gm}, a naive scale-up version of QCD,  was dead due to the excessive 
flavor-changing neutral currents (FCNC).   A solution to the FCNC problem was soon suggested by simply assuming the existence of 
 a large anomalous dimension without any concrete dynamics and concrete  
 value of the anomalous dimension~\cite{Holdom:1981rm}.  
It  was the WTC~\cite{Yamawaki:1985zg,Bando:1986bg} that gave a concrete dynamics,  ladder Schwinger-Dyson (LSD) equation with
 {\it non-running} ({\it scale invariant/conformal}) gauge coupling, $\alpha(p) \equiv \alpha$, 
 giving rise to a concrete value of the anomalous dimension, $\gamma_m = 1$ at criticality $\alpha=\alpha_c$. 
Actually,  once the 
mass $m_F$ of the techni-fermion ($F$) is dynamically generated, 
the coupling becomes  {\it running slowly (``walking'') }  a la Miransky~\cite{Miransky:1984ef}, 
with the {\it nonperturbative} beta function $\beta(\alpha) \sim -  (\alpha/\alpha_c - 1)^{3/2}\quad (\alpha>\alpha_c) $. 
 (Subsequently, a similar FCNC solution based on the ({\it perturbatively}) walking coupling  was  discussed without 
 concept of anomalous dimension~\cite{Akiba:1985rr}.)~\footnote{
Another problem of the TC as a QCD scale-up is the electroweak constraints, so-called $S,T,U$  parameters. 
This may also be improved in the WTC~\cite{Appelquist:1991is,Harada:2005ru}.  
Even if  WTC in isolation cannot overcome this problem, there still exist a possibility that the problem may be 
resolved in the combined dynamical system including the SM fermion mass generation such as the extended TC 
(ETC) dynamics~\cite{Dimopoulos:1979es}, 
in much the same way as the solution (``ideal fermion delocalization'')~\cite{Cacciapaglia:2004rb} in the Higgsless models which simultaneously
adjust S and T parameters by incorporating the SM fermion mass profile. 
  }

In view of the approximate scale symmetry reflected by the {\it nonperturbative} walking coupling
associated with the dynamical generation of  $m_F$,  the WTC predicted the TD~\cite{Yamawaki:1985zg,Bando:1986bg},  
a light scalar $\bar F F$ composite  as a 
pNGB associated with the spontaneous breaking of the 
approximate scale symmetry triggered by the dynamical generation of $m_F$ or the techni-fermion condensate.  
This is in sharp contrast to the TC as a simple scale-up of the QCD 
where the coupling is {\it running already at perturbative level}  with the scale symmetry badly broken at $\Lambda_{\rm QCD}$,
so that  there is  no scale symmetry to be broken by the dynamical fermion mass generation 
and hence no scalar spectrum lighter than the typical hadronic scale, say the rho meson~\footnote{
There might exist a light scalar, so-called $\sigma$ resonance,  in the real-life QCD, which however may not be a two-body composite of $\bar q q$ which is an analogue of 
TD  but may be
mainly a four-body composite. Such a  situation of the real-life QCD is an accidental consequence of  specific values  $N_c=N_f=3$ and $m_u\simeq m_d \ll m_s$.  See e.g.,
\cite{Black:1999yz}. }. 
The mass generation  due to such a scale invariant (conformal) dynamics takes the form of essential-singularity scaling, Miransky scaling~\cite{Miransky:1984ef}, and 
can be  characterized by the ``{\it conformal phase transition}''~\cite{Miransky:1996pd}.

It should be noted here that even if the gauge coupling  is non-running 
(scale invariant)  in the {\it perturbative} sense as in the case of the LSD equation, 
the scale symmetry is actually {\it broken explicitly 
for  the very reason of the spontaneous breaking itself}, namely the dynamical generation of the techni-fermion mass 
$m_F$, which 
is  responsible for the {\it nonperturbative} 
running of the coupling ({\it nonperturbative} scale anomaly) as mentioned above. 
Accordingly, the TD cannot be massless even if we use a {\it perturbatively} non-running (scale invariant) coupling 
in the LSD equation. 
Actually, various old calculations~\cite{Miransky:1989qc, Bardeen:1985sm} imply   
$M_{\rm TD} = {\cal O} (m_F)$, still smaller, 
though not extremely smaller,  than masses of other techni-hadrons.

Modern version of WTC~\cite{Lane:1991qh,Appelquist:1996dq, Miransky:1996pd} is based on the two-loop running coupling 
with the Caswell-Banks-Zaks infrared fixed point (CBZ-IRFP)~\cite{Caswell:1974gg},
 instead of the non-running one, in the improved LSD equation.  
There also exists an intrinsic scale $\Lambda_{\rm TC}$ analogous to $\Lambda_{\rm QCD}$,
 which breaks the scale symmetry already at two-loop perturbative level for the ultraviolet region $p>\Lambda_{\rm TC}$ 
(taken to be $ >\Lambda_{\rm ETC} >10^3 \, {\rm TeV}$) where the coupling runs in the same way as in QCD. 
 However this {\it perturbative} scale-symmetry-breaking scale $\Lambda_{\rm TC}$ is irrelevant to the dynamical mass $m_F$ and so is the TD mass $M_{\rm TD}$, both can be $\ll \Lambda_{\rm TC}$ thanks to the CBZ-IRFP.

Indeed, it has recently been argued \cite{Yamawaki:2009vb,
Vecchi:2010aj} and explicitly 
shown~\cite{Hashimoto:2010nw}  in the case of the two-loop perturbative coupling  that we have essentially the same conclusion as the non-running case in the above, 
$M_{\rm TD} = {\cal O} (m_F) \, (\ll \Lambda_{\rm TC})$, in sharp contrast to the recent claim on much smaller mass 
$M_{\rm TD} \ll {\cal O} (m_F)$~\cite{Appelquist:2010gy}~\footnote{
We here exclude  the ``decoupled TD'' scenario~\cite{Haba:2010hu,Hashimoto:2010nw} 
with the Yukawa coupling  
$\sim m_F/ F_{\rm TD} \to 0 $ as $m_F/\Lambda_{\rm TC} \to 0$, which 
is irrelevant to LHC experiments, although it might be relevant to dark matter~\cite{Hashimoto:2010nw,Choi:2011fy}. 
}. 
Namely, the dynamical generation of $m_F$ triggers  spontaneous breaking of the scale symmetry 
which is also {\it explicitly broken  by $m_F$}  through  the {\it  nonperturbative running} coupling 
({\it nonperturbative scale anomaly}) again a la Miransky.  
At any rate the approximate scale symmetry is crucial for 
the mass of TD to be lighter than other techni-hadrons like techni-rho meson 
with the mass  $M_{\rho}$, 
 $( \Lambda_{\rm TC} \gg)\,   M_{\rho} > {\cal O}({\rm TeV}) >M_{\rm TD} = {\cal O} (m_F)$. 
More concretely, it was noted~\cite{Yamawaki:2009vb} that several earlier  nonperturbative calculations of the scalar mass 
in %the 
different contexts can be interpreted as those 
for the estimate of TD mass~\cite{Shuto:1989te,Harada:2003dc}: 
$M_{\rm TD} \simeq \sqrt{2} m_F$,  which would suggest the TD mass as low as 
\begin{equation}
M_{\rm TD} \simeq  600\, {\rm GeV}
\label{600}
\end{equation}
  (one-family model), well within the reach of LHC searches. 
This is also consistent with the recent holographic estimate~\cite{Haba:2010hu} and others~\cite{Kutasov:2011fr}.
(The above estimate  would suggest  $M_{\rm TD} \simeq 1 \, {\rm TeV}$ for the one-doublet model, 
barely within  the LHC search region)~\footnote{ 
The TD with such a mass region gives substantial negative logarithmic contribution to the T parameter 
in a manner similar to the SM heavy Higgs. 
This could be an extra bonus, since it can in principle be compensated by 
the troublesome positive contributions from the techni-fermion dynamics~\cite{hep-ph/0101342}. }.

In this paper, we explore the characteristic signature of TD at the LHC, calculating the relevant decay widths 
and production cross sections in comparison with those of the SM Higgs. 
To set definite benchmarks, we employ typical models of TC~\cite{Farhi:1980xs}
such as the one-doublet model (1DM)~\footnote{
The 1DM in the usual sense is not walking. We here use ``1DM'' as a modified model (``partially gauged model'')~\cite{Christensen:2005cb} 
which, besides one doublet techni-fermions  with EW  charges as in the usual 1DM, 
 has  dummy techni-fermions without EW charges which only contribute to
the walking behavior of TC dynamics. Actually, such dummy techni-fermions are needed 
even for 1FM with $N_{\rm TC}=3$ since in this case $N_{\rm TF} = 4 N_{\rm TC} = 12 > 8$. } 
and one-family model (1FM) in the light of WTC.

The TD couplings to the SM particles are derived based on nonlinear realization of both 
scale and chiral symmetries, which becomes  highly nontrivial since the scale symmetry, in contrast to the chiral symmetry, is broken explicitly as well as 
spontaneously due to the dynamical mass generation of the
techni-fermion $m_F$, leading to the nonperturbative scale anomaly as mentioned above. There is no limit where TD becomes exactly massless.
Thus the nonlinear realization of the scale symmetry must  
properly include the explicit breaking at the same time as the spontaneous breaking, which we 
will do via spurion field method, the method familiar in the chiral perturbation theory incorporating the current quark mass term~\cite{Gas:84}.
Then the TD couplings 
are given as functions of the techni-fermion mass $m_F$, the TD decay constant $F_{\rm TD}$ (or TD Yukawa coupling)
 and the TD mass $M_{\rm TD}$ up to the number of TC $N_{\rm TC}$.  
Note that $F_{\rm TD}$ may be written in terms of  $M_{\rm TD}$ and $m_F$ through the the partially conserved dilatation current (PCDC) 
for the trace anomaly (nonperturbative scale anomaly) reflecting the spontaneous and explicit breaking of 
the scale symmetry due to dynamical generation of $m_F$,  and $m_F$ may be written in terms of 
the weak scale $v_{\rm EW} =246 \,{\rm GeV}$ through the Pagels-Stokar (PS) formula~\cite{Pagels:1979hd}. 
Hence we can estimate all the quantities only in terms of the TD mass $M_{\rm TD}$ 
as a free parameter which we take in a wide region $200 \,{\rm GeV} <  M_{\rm TD} < 1000$ GeV 
around the reference value 600 GeV 
in Eq.(\ref{600}) suggested by the various calculations~\cite{Yamawaki:2009vb,Haba:2010hu}. 
In order to do a concrete estimate we adopt a 
recent result of 
the nonperturbative scale anomaly~\cite{Hashimoto:2010nw}  
based on the LSD analysis  with 
the two-loop beta function of large $N_f$ QCD having the CBZ-IRFP~\cite{Caswell:1974gg}. 
The result is not qualitatively changed (see the discussion in the last section) 
if we employ the non-running coupling in the LSD as in Refs.~\cite{Yamawaki:1985zg,Bando:1986bg}.

We then explicitly evaluate the 
LHC production cross sections of TD at $\sqrt{s}=7$ TeV, 
$\sigma_{\rm TD}$, times the TD branching ratios,  for 
the TD mass range taken  
within the LHC search region 
$200 \,{\rm GeV} <  M_{\rm TD} < 1000$ GeV. 
We find that the TD signatures are 
quite different from those of the SM Higgs: 
 For 1DMs all those cross sections get suppressed compared to the corresponding quantities for the SM Higgs 
due to the suppression of the gluon 
fusion cross section coming from the large suppression of TD couplings, 
while for 1FMs they get highly enhanced since the production cross section has huge extra contributions from 
the extra colored fermions (techni-quarks). 
As a check of consistency with the current LHC data, 
we compare the cross section $\sigma_{\rm TD} \times BR({\rm TD} \to WW)$ 
normalized to the corresponding quantity for the SM Higgs 
with the recent bounds from the ATLAS~\cite{ATLAS} and CMS~\cite{CMS} experiments~\footnote{
As far as the current Higgs research mass region up to 600 GeV is concerned, 
the narrow width approximation can be applied to TD even in the IFMs 
as well as the SM Higgs, since the TD with mass up to 600 GeV turns out to have 
an almost identical size of total width compared to the SM Higgs.  The narrow width 
approximation is much better in the case of 1DMs which have highly suppressed couplings. 
}.  
 It turns out that in the case of 1DMs the signature is consistent with the current experimental data 
over the whole region we study thanks to the large suppression of TD couplings, and by the same token 
the signal is too tiny to be visible through this channel at LHC. 
As for the 1FMs, on the other hand,  the TD mass is constrained to be excluded up till 
$M_{\rm TD} \simeq 600$ GeV, 
which, however, would imply occurrence of somewhat large excess at $600 \, {\rm GeV} \lesssim M_{\rm TD} <$ 1000 GeV 
in this channel to be seen in the future experiments. 
We further calculate the cross section $\sigma_{\rm TD} \times BR({\rm TD} \to \gamma\gamma)$ 
and predict it to be $\sim 0.10-1.0$ fb at $\sqrt{s}=7$ TeV for the TD mass around 600 GeV in the typical 1FMs. 
This cross section is comparable with the golden mode of the SM Higgs $pp \to h_{\rm SM} \to ZZ \to l^+l^-l^+l^-$, 
and hence is large enough for the TD to be discovered 
within the first few year's run at the LHC.

\section{The TD couplings and decay widths}

 \begin{figure}%[h]
\begin{center}
   \includegraphics[scale=0.7]{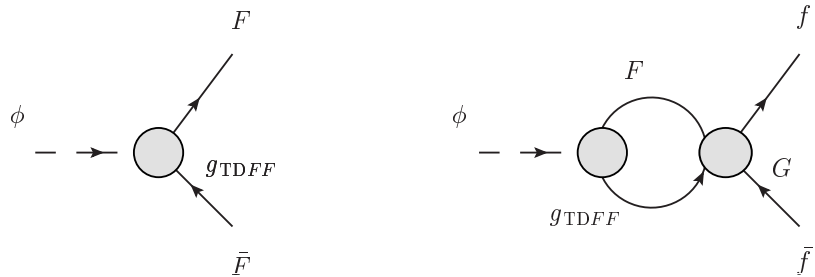}
\caption{ 
The TD Yukawa couplings $g_{{\rm TD} FF}$ (left panel) 
and $g_{{\rm TD}ff}$ (right panel) to techni-fermions ($F$) and SM fermions ($f$). 
The blob denoted as $G$ in the 
right panel corresponds to an ETC-induced four fermi vertex $G \bar{F}F ff$. 
\label{yukawa:fig}
}
\end{center} 
 \end{figure}

 In this section we shall derive the TD couplings to the SM particles and address their forms.    
We focus on the couplings to $WW, ZZ, gg, \gamma\gamma$ and $t \bar{t}$ to make 
the explicit comparison with those of the SM Higgs.

The TD Yukawa couplings to the techni-fermion $g_{{\rm TD}FF}$ and those to the SM fermions $g_{{\rm TD}ff}$ 
were actually derived long time ago~\cite{Bando:1986bg} through 
the Ward-Takahashi identity for dilatation current coupled with TD, 
%They %are obtained by evaluating 
corresponding  to the diagrams depicted in Fig.~\ref{yukawa:fig}:
%through 
%the Ward-Takahashi identity for dilatation current coupled with TD: 
\begin{equation}
g_{{\rm TD}FF} = \frac{(3-\gamma_m)m_{F}}{F_{\rm TD}}\,, \quad g_{{\rm TD}ff} 
= \frac{(3-\gamma_m) m_f}{F_{\rm TD}}\,,
\label{Yukawa}
\end{equation}
where $(3-\gamma_m)$ denotes the scale dimension of techni-fermion bilinear operator 
$\bar{F}F$, which is  $\simeq 2$ for the anomalous dimension $\gamma_m \simeq 1$ in WTC and 
we have assumed an ETC-induced four-fermi interaction such as $G \bar{F}F ff$ which gives the $f$-fermion mass 
$m_f=-G \langle \bar{F}F \rangle$~\footnote{
The top mass is hardly reproduced by the WTC with anomalous dimension $\gamma_m \simeq 1$. 
It may require  other dynamics such as the top quark condensate~\cite{Miransky:1988xi}. 
However, it was found~\cite{Miransky:1988gk} that  if we include additional four-fermion interactions like strong ETC, 
the anomalous dimension becomes much larger $1<\gamma_m<2$,  
which can boost the ETC-origin mass to arbitrarily large up till  the techni-fermion mass scale (``strong ETC model'').  
Subsequently the same effects  were also noted  without concept of the anomalous dimension~\cite{Matumoto:1989hf}. }.

Here  we work on nonlinear realization of 
both the scale and chiral symmetries 
to derive a nonlinear Lagrangian which directly yields 
the various TD couplings to the SM particles. 
Actually, the resultant Lagrangian {\it should not be invariant}  under the scale symmetry 
because it is broken by techni-fermion mass generation explicitly as well as spontaneously 
as was emphasized above. 
We therefore incorporate such an 
inherent explicit-breaking effects (nonperturbative scale anomaly) arising from the dynamical mass generation itself
by %explicitly 
introducing a spurion field in a familiar manner~\cite{Gas:84}, 
which will provide us with a direct way to read appropriate couplings %forms
of TD.

 We begin by introducing a dynamical variable $\Phi$ which reflects the scale transformation property of  
the techni-fermion bilinear $\bar{F}F$ ($\Phi \approx \frac{\bar{F}F}{\langle \bar{F}F  \rangle}$), 
analogously to the usual chiral field $U$ reflecting the chiral transformation property of $\bar{q}_Lq_R$, 
so that $\Phi$ transforms under the scale symmetry as 
\begin{equation} 
  \delta \Phi =  \left(  3-\gamma_m  +  x^\nu \partial_\nu \right) \Phi
\,. \label{trans:Phi}
\end{equation}
We parametrize this $\Phi$ with the TD field $\phi$ and the decay constant $F_{\rm TD}$ as 
\begin{equation} 
\Phi=e^{(3-\gamma_m)\phi/F_{\rm TD}} 
\,, 
\end{equation} 
where $F_{\rm TD}$ is defined as 
\begin{equation}
  \langle0 | D_\mu(0)  | \phi(p) \rangle = - i p_\mu F_{\rm TD} 
\,, 
\end{equation}
with $D_\mu$ being the dilatation current composed {\it only of the TC sector fields}.   
 From Eq.(\ref{trans:Phi}) it follows that the TD field $\phi$ transforms nonlinearly under the scale symmetry: 
\begin{equation} 
  \delta \phi = F_{\rm TD} + x^\nu\partial_\nu \phi 
  \,. 
\end{equation}

To incorporate the 
explicit-breaking effects due to the nonperturbative scale anomaly of the TC sector, we introduce 
a spurion field $S$ which transforms under the scale symmetry with the scale dimension 1: 
\begin{equation} 
  \delta S = \left( 1 +  x^\nu \partial_\nu  \right) S  
\,. 
\end{equation}
Its vacuum expectation value $\langle S \rangle = 1$ thus breaks the scale symmetry explicitly.

We further introduce the usual chiral field $U= e^{2i\pi/v_{\rm EW}}$, with $\pi$ being the NGB fields for the spontaneous chiral symmetry breaking, and consider only the would-be NGBs eaten by $W$ and $Z$ bosons for simplicity~\footnote{
Here we have ignored terms involving techni-pions not eaten by $W$ and $Z$ bosons
 which would appear in models such as 1FM. 
Even if we incorporate them, however,  
the forms of TD couplings to the SM particles given here  
will be intact. }. 
This $U$ and $\pi$ should have scale dimension 0 such that $\pi$ transform  
linearly under the scale symmetry.

 With these at hand, we can write down a nonlinear Lagrangian {\it invariant} under 
the EW and scale symmetries {\it including the spurion field $S$}. 
The Lagrangian is constructed so as to reproduce the appropriate scale anomaly terms coupled to TD 
generated in the underlying WTC when $\langle S \rangle = 1$ is taken.  
It turns out that the Lagrangian including the TD couplings to the SM particles at the leading order 
takes the form~\footnote{
In Ref.~\cite{Clark:1986gx} a similar nonlinear realization was discussed, but it does not take account of  
those inherent explicit-breaking effects arising from the dynamical mass generation. Note that the scale-transformation property of
$(\Phi S^{\gamma_m-2}) \sim e^{\phi/F_{\rm TD}}$, which is the same as the nonlinear base 
used in  Ref.~\cite{Clark:1986gx} unless taking $\langle S \rangle = 1$.} 
 (The explicit proof of the Wess-Zumino type consistency with the scale anomaly is to be given in another publication.)  
\begin{eqnarray} 
  {\cal L} 
&=& \frac{v_{\rm EW}^2}{4} (\Phi S^{\gamma_m-2})^2 
{\rm tr}[D_\mu U^\dagger D^\mu U] 
- ( \Phi S^{\gamma_m-2}) \, \sum_f \left( \bar{f}_L U  
\left( 
\begin{array}{cc}  
  m_f^u & 0 \\ 
  0 & m_f^d 
\end{array}
\right) 
f_R +  {\rm h.c.} \right)
  \nonumber \\
&& 
-  (\Phi S^{\gamma_m-3}) 
 \left( \frac{\beta_F(\alpha_s)}{2 \alpha_s} {\rm tr}[G_{\mu\nu}^2] + \frac{\beta_F(\alpha_{\rm EM})}{4 \alpha_{\rm EM}} F_{\mu\nu}^2 
  \right) 
  \,, \label{L}
\end{eqnarray} 
where $D_\mu U =\partial_\mu U - i g_W W_\mu^a \frac{\sigma^a}{2} U + i g_Y U B_\mu \frac{\sigma^3}{2}$; 
$W_\mu^a$ ($a=1,2,3$) and $B_\mu$ are the $SU(2)_W$ and $U(1)_Y$ gauge fields with the gauge couplings $g_W$ and $g_Y$; 
$\sigma^a$ denotes Pauli matrices; 
$G_{\mu\nu}$ and $F_{\mu\nu}$ are field strengths for the QCD gluon and electromagnetic (EM) gauge (photon) fields 
with the gauge couplings $\alpha_{s,{\rm EM}}=g_{s,{\rm EM}}^2/(4\pi)$, respectively; $\beta_F$ denotes beta function including 
only the techni-fermion loop contributions; $f_{L,R}=(f_{L,R}^u, f_{L,R}^d)^T$ stand for the $SU(2)_{L,R}$ doublets 
 with $m_f^{u,d}$ being their masses.  Note that $\beta_F(\alpha_s)=0$ in the case of 1DM, while $\beta_F(\alpha_{\rm EM}) \neq 0$ 
 because of the presence of techni-fermions having the EM charges.

\begin{figure}%[h]
\begin{center}
   \includegraphics[scale=0.6]{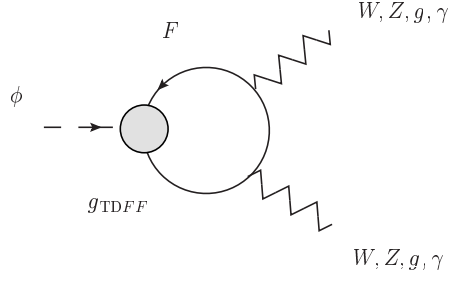}
\caption{ 
The TD couplings to $WW,ZZ,gg,\gamma\gamma$ induced from techni-fermion loops. 
\label{gauge:fig}
}
\end{center} 
 \end{figure}

 Taking  $\langle S \rangle = 1$, from Eq.(\ref{L}) 
we readily find the TD couplings to $WW$, $ZZ$ and $\bar{f}f$: 
\begin{eqnarray} 
  {\cal L}_{{\rm TD}WW/ZZ} 
&=& 
g_{{\rm TD}WW} \, \phi W_\mu^+ W^{\mu -} + \frac{1}{2} g_{{\rm TD}ZZ} \, \phi Z_\mu Z^\mu   
\,,   \\ 
{\cal L}_{{\rm TD}ff} 
&=&~- g_{{\rm TD}ff} \, \phi \bar{f}f 
\,, \\ 
g_{{\rm TD}WW/ZZ} 
&=&  \frac{2 (3-\gamma_m) m_{W/Z}^2}{F_{\rm TD}} 
\,, \qquad  
g_{{\rm TD}ff} 
=  \frac{ (3-\gamma_m) m_f}{F_{\rm TD}}
\,, \label{TD-WWZZ} 
\end{eqnarray}
and the couplings to $gg$ and $\gamma\gamma$, 
\begin{eqnarray} 
  {\cal L}_{{\rm TD} gg/\gamma\gamma} 
&=&  
   - g_{{\rm TD}gg} \,  \phi {\rm tr}[G_{\mu\nu}^2] 
   - g_{{\rm TD}\gamma\gamma} \,  \phi F_{\mu\nu}^2 
  \,, \\ 
  g_{{\rm TD} gg} &=& \frac{ (3-\gamma_m)}{F_{\rm TD}} \frac{\beta_F(\alpha_s)}{2 \alpha_s} 
  \,, \qquad 
    g_{{\rm TD} \gamma\gamma} =
\frac{ (3-\gamma_m)}{F_{\rm TD}} \frac{\beta_F(\alpha_{\rm EM})}{4 \alpha_{\rm EM}}
\,. \label{TD-gg2gamma:2}
\end{eqnarray} 
The same couplings as the above can actually be obtained by 
directly evaluating diagrams in Fig.~\ref{gauge:fig} and the right panel of Fig.~\ref{yukawa:fig},  
hence the results for $ g_{{\rm TD}ff} $ and $g_{{\rm TD}WW/ZZ} $ 
are identical to Eq.(\ref{Yukawa}) and that of Ref.~\cite{Hashimoto:2011cw}, respectively. 
The couplings in Eqs.(\ref{TD-WWZZ}) and (\ref{TD-gg2gamma:2}) 
are compared with the SM Higgs couplings 
$g_{{\rm TD}WW/ZZ} = \frac{2 m_{W/Z}^2}{v_{\rm EW}}$, 
$g_{{\rm TD}ff} = \frac{m_f}{v_{\rm EW}}$, 
$ g_{h_{\rm SM} gg}= \frac{1}{v_{\rm EW}} \frac{\beta(\alpha_s)}{ 2\alpha_s} $,  
and 
$ g_{h_{\rm SM} \gamma\gamma}= \frac{1}{v_{\rm EW}} \frac{\beta(\alpha_{\rm EM})}{4 \alpha_{\rm EM}} $,  
which indeed implies a simple replacement, 
$1/v_{\rm EW} \to (3-\gamma_m)/F_{\rm TD}\, ( \simeq 2/F_{\rm TD}$ for  $\gamma_m \simeq 1$), 
between the SM Higgs and TD couplings. 
 (Note that $1/v_{\rm EW} \ne (3-\gamma_m)/F_{\rm TD}$, with the value of $F_{\rm TD}$ being related to $v_{\rm EW}$ in a highly model dependent way.)  
The essential discrepancy in coupling forms thus arises only as the overall coupling strengths 
set by the TD decay constant $F_{\rm TD}$, in place of the EW scale $v_{\rm EW}\simeq 246$ GeV.

Once the TD couplings are obtained, 
we can calculate the TD decay widths $\Gamma({\rm TD} \to X)$ $(X=WW,ZZ,gg, \gamma\gamma,t\bar{t})$ 
evaluating the amplitudes at the leading order of perturbation. 
Actually, the $gg$ and $\gamma\gamma$ couplings in Eq.(\ref{TD-gg2gamma:2}) 
are not sufficient for evaluating the ${\rm TD} \to gg$ and $\gamma\gamma$ decays  
since these terms arise as higher dimensional (derivative) operators 
more sensitive to ultraviolet contributions of ${\cal O}(m_F)$ than 
the TD-$WW, ZZ$ and $t\bar{t}$ coupling terms in Eq.(\ref{TD-WWZZ}) which are of the lowest order. 
Therefore we shall straightforwardly evaluate techni-fermion loop contributions to 
the ${\rm TD} \to gg$ and $\gamma\gamma$ decay widths, instead of using 
the operator coupling forms in Eq.(\ref{TD-gg2gamma:2}).

We assume that all the techni-fermions belong to the fundamental representation in the TC gauge group 
of $SU(N_{\rm TC})$ and have the flavor-independent mass $m_F$ 
defined as usual $\Sigma(p=m_F)=m_F$ in the LSD equation analysis, 
where $\Sigma(p)$ is the mass function of the techni-fermions. 
Here we use the constant mass function $\Sigma(p) \equiv m_F$ for 
calculating the amplitudes for simplicity. 
Even if we use the momentum-dependent mass function (solution of the LSD) throughout all the calculations, 
the result will not be changed as far as the LHC energy $p \lesssim m_F < {\cal O}({\rm TeV})$ is concerned. 
Also used is the TD Yukawa coupling to techni-fermions 
$g_{{\rm TD}FF}$ given in Eq.(\ref{Yukawa}).

We further add the SM loop contributions to the decay widths $\Gamma({\rm TD} \to gg)$ 
and $\Gamma({\rm TD} \to \gamma\gamma)$, which will be relevant in magnitude as in the case of the SM Higgs, 
although those are of subleading order in terms of TC dynamics.

The decay widths are thus calculated to be 
\begin{eqnarray} 
 \Gamma({\rm TD} \to gg) 
 &=& \frac{\alpha_s^2 M_{\rm TD}^3}{8 \pi^3 F_{\rm TD}^2} 
 \Bigg| 
  \sum_{f=t,b} \tau_f \left[ 1 + \left( 1 - \tau_f \right) f(\tau_f) \right] 
  + N_{\rm TC}
  \sum_{F \, \textrm{with QCD color}} \tau_F \left[ 1 + \left( 1 - \tau_F \right) f(\tau_F) \right] 
  \Bigg|^2 
 \,, \label{TDgg} \\ 
  \Gamma({\rm TD} \to \gamma\gamma) 
 &=& \frac{\alpha_{\rm EM}^2 M_{\rm TD}^3}{64 \pi^3 F_{\rm TD}^2} 
 \Bigg| 
 A_W(\tau_W)+ 
3 \sum_{f=t,b} Q_f^2 A_f(\tau_f) 
  + 
 N_{\rm TC} \sum_{F}  N_c^{(F)} Q_F^2 A_F(\tau_F) 
  \Bigg|^2 
  \,, \label{TD:2gamma}
\end{eqnarray}
 where $\tau_i=4m_i^2/M_{\rm TD}^2$ ($i=W, Z, f,F$); $N_c^{(F)}=3(1)$ for techni-quarks (leptons); $Q_{f(F)}$ denotes 
 EM charge for $f(F)$-fermion, and 
\begin{eqnarray} 
 A_W(\tau_W) &=& -2 \left[ 2 + 3 \tau_W + 3 \tau_W (2-\tau_W) f(\tau_W) \right] 
\,,  \\ 
 A_{f(F)}(\tau_{f(F)}) &=&  2 \tau_{f(F)} \left[ 1 + \left( 1 - \tau_{f(F)} \right) f(\tau_{f(F)}) \right] 
\,, \\ 
  f(\tau_i) 
&=&  
  \Bigg\{ 
  \begin{array}{cc} 
  \left( \sin^{-1} \frac{1}{\sqrt{\tau_i}} \right)^2 & {\rm for} \qquad \tau_i > 1 \nonumber \\ 
  - \frac{1}{4} \left[  \log \left( \frac{1 + \sqrt{1+\tau_i}}{1 - \sqrt{1-\tau_i}} \right) - i \pi \right] 
  & {\rm for} \qquad \tau_i \le 1 
  \end{array}
  \,. 
\end{eqnarray} 
The ${\rm TD} \to gg$ and $\gamma\gamma$ decay widths are thus quite sensitive to 
type of models of WTC, which leads to characteristic signatures of TD highly model-dependent, 
as will be seen later.

As for the formulas for $\Gamma({\rm TD} \to WW/ZZ/t\bar{t})$,    
it turns out that the resultant expressions for those decay widths actually take the same forms as 
the familiar ones for the SM Higgs, say, as listed in Ref.~\cite{Spira:1997dg}: 
\begin{eqnarray} 
\Gamma({\rm TD} \to WW/ZZ) 
&= & \delta_{W(Z)} \, \frac{M_{\rm TD}^3}{8\pi F_{\rm TD}^2} 
\sqrt{1- \tau_{W/Z}} (1 - \tau_{W/Z} + \frac{3}{4} \tau_{W/Z}^2) 
\,, \label{TD:WWZZ} \\ 
\Gamma({\rm TD} \to t\bar{t}) 
&=& 
\frac{3 m_t^2 M_{\rm TD}}{2\pi F_{\rm TD}^2} 
\left( 1 - \tau_t \right)^{3/2}
\,, \label{TD:ttbar}
\end{eqnarray}
where $\delta_{W(Z)} =2(1)$. 
It is easily checked that these formulas are reduced to the SM Higgs ones 
just by replacing  $F_{\rm TD}$ as  $F_{\rm TD} \to 2 v_{\rm EW}$ 
(We have used $(3-\gamma_m) =2$ in the above formulas).

The TD decay widths are thus obtained as functions of the TD mass $M_{\rm TD}$, 
the decay constant $F_{\rm TD}$ and techni-fermion mass $m_F$ 
in addition to the number of TC $N_{\rm TC}$.

\section{The TD LHC signatures at $\sqrt{s}=7$ TeV}

In this section we shall discuss the TD LHC production 
cross sections times the branching ratios. 
The dominant production cross section arises through gluon fusion (GF) and vector boson fusion (VBF) processes similarly to 
the SM Higgs case. 
We thus consider these cross sections times branching ratios normalized to those of the SM Higgs: 
\begin{equation} 
    R_{X} 
    \equiv 
    \frac{ [\sigma_{\rm GF}(pp \to {\rm TD}) + \sigma_{\rm VBF}(pp \to {\rm TD})]}{
[\sigma_{\rm GF}(pp \to h_{\rm SM}) + \sigma_{\rm VBF}(pp \to h_{\rm SM})]} 
\frac{BR({\rm TD} \to X)}{BR(h_{\rm SM} \to X)}
\,, \label{R:0}
\end{equation}
where $X=WW, ZZ, gg,\gamma\gamma$ and $t\bar{t}$. 
The ratios of the production cross sections are related to the ratios of the corresponding 
decay widths as~\cite{Georgi:1977gs} 
\begin{equation} 
    \frac{\sigma_{\rm VBF}(pp \to {\rm TD})}{\sigma_{\rm VBF}(pp \to h_{\rm SM})} 
  = \frac{\Gamma({\rm TD} \to WW)}{\Gamma(h_{\rm SM} \to WW)} 
  = \frac{\Gamma({\rm TD} \to ZZ)}{\Gamma(h_{\rm SM} \to ZZ)} 
   \equiv r_{WW/ZZ} 
    \,, 
  \qquad 
  \frac{\sigma_{\rm GF}(pp \to {\rm TD})}{\sigma_{\rm GF}(pp \to h_{\rm SM})} 
  = \frac{\Gamma({\rm TD} \to gg)}{\Gamma(h_{\rm SM} \to gg)} 
  \equiv r_{gg} 
  \,. 
  \label{r}
\end{equation} 
   Hence we may rewrite Eq.(\ref{R:0}) as  
\begin{equation}
R_{X} 
=
\left( 
\frac{\sigma_{\rm GF}(pp \to h_{\rm SM}) \cdot r_{gg}+\sigma_{\rm VBF}(pp \to h_{\rm SM})\cdot r_{WW/ZZ}}{\sigma_{\rm GF}(pp \to h_{\rm SM}) + \sigma_{\rm VBF}(pp \to h_{\rm SM})}
\right) 
r_{\rm BR}^X
\,, \label{R}
\end{equation}
where 
\begin{equation}
  r_{\rm BR}^X
  = 
  \frac{BR({\rm TD} \to X)}{BR(h_{\rm SM} \to X)}
  \,. \label{rBR}
\end{equation} 
The SM Higgs production cross sections $\sigma_{\rm GF}(pp \to h_{\rm SM})$ and $\sigma_{\rm VBF}(pp \to h_{\rm SM})$ at $\sqrt{s}=7$ TeV 
are read off from Ref.~\cite{CERN}.

For the explicit estimate of $R_X$ in Eq.(\ref{R}),  
we shall consider typical models of WTC (1DM and 1FM) and  
adopt the recent results from the LSD analysis~\cite{Hashimoto:2010nw} to 
specify the values of techni-fermion mass $m_F$ and TD decay constant $F_{\rm TD}$ 
in such a way that the decay widths 
are expressed only in terms of the TD mass $M_{\rm TD}$.

  From Ref.~\cite{Hashimoto:2010nw} we read off the result on $m_F$ obtained through  
  the PS formula for the techni-pion decay constant $F_\pi$ which is related to the EW scale $v_{\rm EW}$ as 
$v_{\rm EW} = \sqrt{N_{\rm D}} F_\pi$ where $N_{\rm D}$ denotes 
the number of EW doublets which equals to half of 
the number of techni-fermions charged under the EW gauge: $N_{\rm  D}=1$ for 1DM and
$N_{\rm D}= 4$ for 1FM. 
Adding dummy techni-fermions~\cite{Christensen:2005cb} which are singlet under the EW gauge, 
the total number of techni-fermions $N_{\rm TF}$ is expressed as 
\begin{equation} 
N_{\rm TF} = (N_{\rm TF})_{\rm EW-singlet} + 2 N_{\rm D} 
\,. 
\end{equation} 
At the criticality where 
  the CBZ-IRFP $\alpha_*$ coincides with the critical coupling $\alpha_c$ for the chiral symmetry breaking, 
  the PS formula goes like 
\begin{equation} 
  \frac{v_{\rm EW}}{m_F} \simeq 0.41 \left( \frac{N_{\rm TC}}{3} \right)^{1/2} \left( \frac{N_{\rm D}}{1}  \right)^{1/2}
  \,. \label{PS}
\end{equation}
   From this we have  
\begin{eqnarray} 
  m_F &\simeq& 600 \, {\rm GeV}  \, \left( \frac{N_{\rm TC}}{3} \right)^{-1/2} \left( \frac{N_{\rm D}}{1}  \right)^{-1/2}
\nonumber \\ 
  &\simeq& 
  \Bigg\{ 
  \begin{array}{cc} 
  735 \, (600)  \, {\rm GeV} &  \qquad  \textrm{for the 1DM ($N_{\rm D}=1$) with $N_{\rm TC}=2(3)$} \\ 
  367 \, (300)  \, {\rm GeV} &  \qquad  \textrm{for the 1FM ($N_{\rm D}=4$) with $N_{\rm TC}=2(3)$} 
  \end{array} 
\,. \label{mF}
\end{eqnarray}

By using the PCDC relation, on the other hand, the TD decay constant $F_{\rm TD}$ and TD mass $M_{\rm TD}$ 
are related with vacuum energy $V$ through the trace anomaly (nonperturbative scale anomaly induced by the dynamical generation of $m_F$) 
as follows: 
\begin{equation} 
  F_{\rm TD}^2 M_{\rm TD}^2 = - d_\theta \langle \theta_\mu^\mu\rangle= - 16 V 
  \,, 
  \label{scaleanomaly}
\end{equation}
where $d_\theta (=4) $ is the scale dimension of trace of the energy-momentum tensor $\theta_\mu^\mu$. 
Here the vacuum energy $V=d_\theta \langle \theta_\mu^\mu  \rangle/16$ only includes contributions 
from the nonperturbative scale anomaly, defined by subtracting contributions $\langle \theta_\mu^\mu \rangle_{\rm perturbation}$ of ${\cal O}(\Lambda_{\rm TC}^4)$ 
from the perturbative running of the gauge coupling $\alpha$, such as 
$\langle \theta_\mu^\mu  \rangle  - \langle \theta_\mu^\mu \rangle_{\rm perturbation}$. 
  To the vacuum energy $V$ the LSD analysis in Ref.~\cite{Hashimoto:2010nw} gives  
\begin{equation} 
 - 4V \simeq 0.76 \left( \frac{N_{\rm TF} N_{\rm TC}}{2\pi^2} \right) m_F^4 
 \,, 
\end{equation}  
at the criticality, and hence 
\begin{equation} 
  F_{\rm TD}^2 M_{\rm TD}^2 \simeq 3.0 \left( \frac{N_{\rm TF} N_{\rm TC}}{2\pi^2} \right) m_F^4
  \,. \label{PCDC}
\end{equation} 
This relation reflects the appropriate dependences of 
$F_{\rm TD}$ on $N_{\rm TC}$ and $N_{\rm TF}$: 
$F_{\rm TD}$ scales with $N_{\rm TF}$ as well as $N_{\rm TC}$ 
like $F_{\rm TD} \propto \sqrt{N_{\rm TC} N_{\rm TF}}$~\cite{Gusynin:1987em}.  
Note also that the (pole) masses $M_{\rm TD}$ and $m_F$ do not scale with $N_{\rm TC}$ and $N_{\rm TF}$.

 From Eq.(\ref{PCDC}) and Eq.(\ref{mF}) 
we obtain $F_{\rm TD}$ as a function of $M_{\rm TD}$: 
\begin{eqnarray} 
  F_{\rm TD} &\simeq& 1413 \, {\rm GeV} \left( \frac{600\,{\rm GeV}}{M_{\rm TD}} \right) 
  \left( \frac{N_{\rm TF}}{4 N_{\rm TC}} \right)^{1/2} \left( \frac{N_{\rm TC}}{3} \right)^{-1/2} \left( \frac{N_{\rm D}}{1}  \right)^{-2}
\nonumber \\ 
    &\simeq& 
  \Bigg\{ 
  \begin{array}{cc} 
  1413 \, (1413)  \, {\rm GeV} \left( \frac{600\,{\rm GeV}}{M_{\rm TD}} \right) &  
\qquad  \textrm{for the 1DM with $N_{\rm TC}=2(3)$, $N_{\rm TF} \simeq 8(12)$} \\ 
  353 \, (353)  \, {\rm GeV} \left( \frac{600\,{\rm GeV}}{M_{\rm TD}} \right) &  
\qquad  \textrm{for the 1FM with $N_{\rm TC}=2(3), N_{\rm TF} \simeq 8(12)$} 
  \end{array} 
  \,, \label{FTD}
\end{eqnarray}
 where we have used $N_{\rm TF}\simeq 4 N_{\rm TC}$~\cite{Appelquist:1996dq} 
obtained by estimating the critical number of flavors at which we have $\alpha_* \simeq \alpha_c$. 
Note that Eq.(\ref{FTD}) merely shows the reference values of $F_{\rm TD}$, 
not reflecting scaling properties with respect to $N_{\rm TC}$ and $N_{\rm D}(N_{\rm TF})$ 
because the pion decay constant 
$F_\pi(v_{\rm EW})$ $\propto \sqrt{N_{\rm TC}}$ has been fixed through $m_F$ fixed as in Eq.(\ref{PS}).

The values of $F_{\rm TD}$ are somewhat larger than the pion decay constant $F_\pi \simeq 246$ GeV (123 GeV) for 
the 1DM (1FM). 
It turns out that the largeness of $F_{\rm TD}$ essentially comes from the smallness of $M_{\rm TD}$
tied with the existence of the approximate scale invariance: 
To see this more clearly, we shall go back to Eq.(\ref{PCDC}) and express 
$m_F$ in terms of $F_\pi$ through Eq.(\ref{mF}) with $v_{\rm EW}=F_\pi/\sqrt{N_{\rm D}}$, 
and then divide both sides of Eq.(\ref{PCDC}) by $F_\pi^4$. 
We then arrive at 
\begin{eqnarray} 
  \frac{F_{\rm TD}}{F_\pi} &\simeq& 
7.0 \times \left( \frac{F_\pi}{M_{\rm TD}} \right) 
  \sqrt{\frac{N_{\rm TF}}{N_{\rm TC}}} 
\nonumber\\ 
& \simeq & 
    \Bigg\{ 
  \begin{array}{cc} 
  5.7 \, \left( \frac{600\,{\rm GeV}}{M_{\rm TD}} \right) &  
\qquad  \textrm{for the 1DMs with $N_{\rm TF} \simeq 4N_{\rm TC}$ and $F_\pi=246$ GeV} \\ 
  2.8  \, \left( \frac{600\, {\rm GeV}}{M_{\rm TD}} \right) &  
\qquad  \textrm{for the 1FMs with  $N_{\rm TF} \simeq 4 N_{\rm TC}$ and $F_\pi = 123$ GeV} 
  \end{array} 
  \,. \label{FTD-Fpi}
\end{eqnarray}
If we had %would have 
$F_{\rm TD} \sim F_\pi$,  then $M_{\rm TD}$ would have to be $\gtrsim {\cal O}$(TeV) which is as large as 
masses of other techni-hadrons like techni-rho meson, as in the case of QCD ``dilaton" such as $f_0(1370)$ (or $f_0(600)$). 
Thus the existence of the approximate scale invariance allowing the small $M_{\rm TD}$ essentially causes 
the large $F_{\rm TD}$.

The TD Yukawa coupling to the SM fermions normalized to the SM Higgs one is estimated for each model:   
\begin{eqnarray} 
  \frac{g_{{\rm TD} ff}}{g_{h_{\rm SM} ff}} 
  &=& \frac{(3-\gamma_m)v_{\rm EW}}{F_{\rm TD}} \Bigg|_{\rm \gamma_m\simeq 1}
\nonumber \\ 
&\simeq& 
(3-\gamma_m)|_{\gamma_m\simeq1} \times 
  \Bigg\{ 
  \begin{array}{cc} 
  0.18 \,  \left( \frac{M_{\rm TD}}{600\,{\rm GeV}} \right) &  
\qquad  \textrm{for the 1DM with $N_{\rm TF}\simeq 4 N_{\rm TC}$} \\ 
  0.71 \,  \left( \frac{M_{\rm TD}}{600\,{\rm GeV}} \right) &  
\qquad  \textrm{for the 1FM with $N_{\rm TF}\simeq 4 N_{\rm TC}$} 
  \end{array} 
  \,. \label{Yukawa:ratio}
\end{eqnarray}
Thus the Yukawa coupling in the 1DM gets suppressed and so do other TD couplings, mainly due to the smallness of $M_{\rm TD}$ as noted above. 
In the case of 1FM, on the other hand,  its suppression is mild due to the relatively smaller $F_{\pi}$ related to the smaller $F_{\rm TD}$ 
(See Eqs.(\ref{FTD}) and (\ref{FTD-Fpi})), so that the extra factor $(3-\gamma_m) \simeq 2$ 
finally pulls the Yukawa coupling strength up to be comparable to the SM Higgs one. 
Note that were it for $M_{\rm TD} \simeq 1.7$ TeV (430 GeV) in the 1DM (1FM),  we would have $g_{{\rm TD} ff} \simeq g_{h_{\rm SM}ff}$, 
which would imply that the TD signature in the 1DM then would become almost identical to those of the SM Higgs 
(except small corrections to the $\gamma\gamma$ mode coming from extra EM-charged techni-fermions).  
It is not the case, on the other hand, for the 1FM because of the presence of extra techni-quarks yielding 
significant contributions to the GF production cross section as will be seen below.

The TD decay widths in Eqs.(\ref{TDgg})-(\ref{TD:2gamma}) and (\ref{TD:WWZZ})-(\ref{TD:ttbar})  
and the ratio $r_{\rm BR}^{X}$ in Eq.(\ref{rBR}) are thus calculated explicitly 
%to be obtained 
as functions of only the TD mass $M_{\rm TD}$. 
The comparison of the TD branching ratios with those of the SM Higgs is shown in 
Table~\ref{list:branching} for the reference value $M_{\rm TD}=600$ GeV in the 1DM and 1FMs 
with $N_{\rm TC}=2, 3$ and the corresponding values of $m_F$ given in Eq.(\ref{mF}). 
  In the case of 1DMs the TD branching fraction becomes almost identical to that of the SM Higgs 
  since the overall difference between the coupling strengths cancels out in the branching ratios. 
Also for 1FMs the same argument is applicable to the decays to $WW,ZZ$ and $t\bar{t}$ as well, however, 
not to the decays to $gg$ and $\gamma\gamma$, which are rather enhanced 
mainly due to the presence of extra colored (techni-quarks)/EM-charged particles contributing to these decay processes.

\begin{table}%[h]
\begin{center}
\begin{tabular}{|c|c|c|c|c|c|c|c|} 
\hline 
Model & $N_{\rm TC}$ 
& \hspace{10pt} $r^{WW}_{\rm BR}$ \hspace{10pt}  
& \hspace{10pt} $r^{ZZ}_{\rm BR}$ \hspace{10pt} 
& \hspace{10pt} $r^{gg}_{\rm BR}$ \hspace{10pt} 
& \hspace{10pt} $r^{\gamma\gamma}_{\rm BR}$ \hspace{10pt} 
& \hspace{10pt} $r^{t\bar{t}}_{\rm BR}$ \hspace{10pt} 
\\
\hline \hline 
1DM & 
2 
& 1.0   & 1.0 
& 1.0 &  1.0  & 0.92   
\\ 
& 
3 
& 1.0   & 1.0 
& 1.0  & 0.80  & 1.0   
\\ 
 \hline 
 1FM & 
2 
& 1.0   & 0.99 
& 16 & 3.2 & 1.0   
\\ 
& 
3 
& 0.99  &  0.99  
& 44 & 11 & 0.99   
\\ 
 \hline 
\end{tabular} 
\caption{ 
The TD branching ratios at $M_{\rm TD}=600$ GeV normalized to the corresponding quantities for the SM Higgs, $r_{\rm BR}^X$ 
($X=WW,ZZ,gg,\gamma\gamma,t\bar{t}$), defined in Eq.(\ref{r}). }
\label{list:branching}
\end{center}
\end{table}

We next pay attention to the production cross sections.  
As seen from Eq.(\ref{R}), the rate of the production cross section to the SM Higgs is determined 
by the amounts of $r_{gg}$ and $r_{WW/ZZ}$ defined in Eq.(\ref{r}) 
quite sensitive to the TD Yukawa couplings. 
The case of 1FM makes the situation most sensitive because of the 
presence of techni-quarks with the number of 2$N_{\rm TC}$ 
which in general enhances the GF cross section $r_{gg}$~\footnote{
A similar enhancement of GF process in the case of 1FM was discussed in Ref.~\cite{Hashimoto:2011ma}. }. 
 For 1DMs, in contrast, it is not the case because of the absence of techni-quarks 
and the suppression of Yukawa coupling 
coming from somewhat larger $F_{\rm TD}$ sensitively reflecting the smallness of $M_{\rm TD}$ 
(See Eq.(\ref{FTD}) and discussion below Eq.(\ref{FTD-Fpi})).

In Table~\ref{list} we list the reference values of $r_{gg}$ and $r_{WW/ZZ}$ for 
$M_{\rm TD}=600$ GeV in the case of 1DM and 1FMs with $N_{\rm TC}=2, 3$ 
together with the values of $g_{{\rm TD} ff}/g_{h_{\rm SM}ff}$ in Eq.(\ref{Yukawa:ratio}). 
The VBF cross section in the case of 1FMs is thus almost of the same order of magnitude as that of the SM Higgs 
because of the almost identical Yukawa coupling strength,  
while the GF cross section gets highly enhanced by a factor of ${\cal O}(10)-{\cal O}(10^2)$ 
depending on %ent of 
$N_{\rm TC}$ due to the techni-quark contributions.  
  For 1DMs without such an enhancement, on the other hand,  both cross sections get suppressed simply due to 
  the suppressed Yukawa couplings.

\begin{table}%[h]
\begin{center}
\begin{tabular}{|c|c|c|c|c|} 
\hline 
Model &
$N_{\rm TC}$ & 
\hspace{8pt} $\frac{g_{{\rm TD} ff}}{g_{h_{\rm SM}ff}}=\frac{2v_{\rm EW}}{F_{\rm TD}}$ 
\hspace{8pt}  
& 
\hspace{8pt} 
$r_{gg}=\frac{\sigma_{\rm GF}^{\rm TD}}{\sigma_{\rm GF}^{h_{\rm SM}}}$ 
\hspace{8pt} 
&  
\hspace{8pt} 
$r_{WW/ZZ}=\frac{\sigma_{\rm VBF}^{\rm TD}}{\sigma_{\rm VBF}^{h_{\rm SM}}}$ 
\hspace{8pt} 
\\
\hline \hline 
1DM & 
2& 0.35 & 0.12 & 0.12  \\ 
& 
3 & 0.35 & 0.12 & 0.12 \\ 
\hline  
1FM & 
2 
& 1.4 & 31 & 1.9  \\ 
& 
3 
& 1.4 & 87 & 1.9  \\ 
 \hline   
\end{tabular} 
\caption{ 
Values of $r_{gg}$ and $r_{WW/ZZ}$ for $N_{\rm TC}=2, 3$ 
in the case of 1FM and 1DMs with $M_{\rm TD}=600$ GeV fixed.  
Also shown are values of the ratio of the Yukawa coupling $g_{{\rm TD} ff}/g_{h_{\rm SM}ff}$. }
\label{list}
\end{center}
\end{table}

Now we calculate the value of $R_X$ in Eq.(\ref{R}) for each channel 
to address more clearly how we can distinguish the TD signatures from those of the SM Higgs at the LHC. 
The results are listed in Table~\ref{list:signals} with $M_{\rm TD}=600$ GeV fixed.

 From Table~\ref{list:signals} we see that in the case of 1DMs 
 all the signatures are suppressed to be one order of magnitude smaller than   
 the corresponding quantities for the SM Higgs 
due to the large suppression of the production cross sections coming from the suppression of 
Yukawa coupling (See Eq.(\ref{Yukawa:ratio}) or Table~\ref{list}). 
It is interesting to note, in particular, that the $WW$ and $ZZ$ modes get suppressed 
in contrast to the SM Higgs case, to be distinguishable from those of the SM Higgs at the LHC.

In the case of 1FMs, on the other hand, all the signals get enhanced due to 
the large GF production cross section highly enhanced by the extra colored-techni-quark contributions (See Table~\ref{list}).  
 This enhancement gets more operative for the $gg$ and $\gamma\gamma$ modes to result in a  
gigantic enhancement mainly because of their highly enhanced branching ratios (See Table~\ref{list:branching}). 
Note that the LHC cross section for the $\gamma\gamma$ mode is quite small 
for the SM Higgs with the mass around 600 GeV, which is about $10^{-4}-10^{-3}$ fb. 
Besides the enhanced $WW$ and $ZZ$ modes, therefore,  
the $\gamma\gamma$ mode will be a characteristic signature of TD clearly distinguishable from 
the SM Higgs to be visible at the LHC, as will explicitly be shown below. 
The enhancement in the $t\bar{t}$ mode may also be a certain TD signature (See the discussion in the last section),  
while the $gg$ mode is inaccessible for the TD searches because of its huge background.

\begin{table}%[h]
\begin{center}
\begin{tabular}{|c|c|c|c|c|c|c|c|} 
\hline 
Model & $N_{\rm TC}$ 
& \hspace{10pt} $R_{WW}$ \hspace{10pt}  
& \hspace{10pt} $R_{ZZ}$ \hspace{10pt} 
& \hspace{10pt} $R_{gg}$ \hspace{10pt} 
& \hspace{10pt} $R_{\gamma\gamma}$ \hspace{10pt} 
& \hspace{10pt} $R_{t\bar{t}}$ \hspace{10pt} 
\\
\hline \hline 
1DM & 
2 
& 0.12  & 0.12
& 0.12 & 0.095 & 0.12  
\\ 
& 
3 
& 0.12  & 0.12
& 0.12 & 0.097  & 0.12  
\\ 
 \hline 
 1FM & 
2 
& 26  & 26 
& 414 & 85 & 26  
\\ 
& 
3 
& 73  &  73 
& 3300 & 840 & 73  
\\ 
 \hline 
\end{tabular} 
\caption{ 
The TD signatures at $M_{\rm TD}=600$ GeV normalized to the corresponding quantities for the SM Higgs, $R_X$ 
($X=WW,ZZ,gg,\gamma\gamma,t\bar{t}$) defined in Eq.(\ref{R}). }
\label{list:signals}
\end{center}
\end{table}

\begin{figure}%[h]
\begin{center}
 \includegraphics[width=8.5cm]{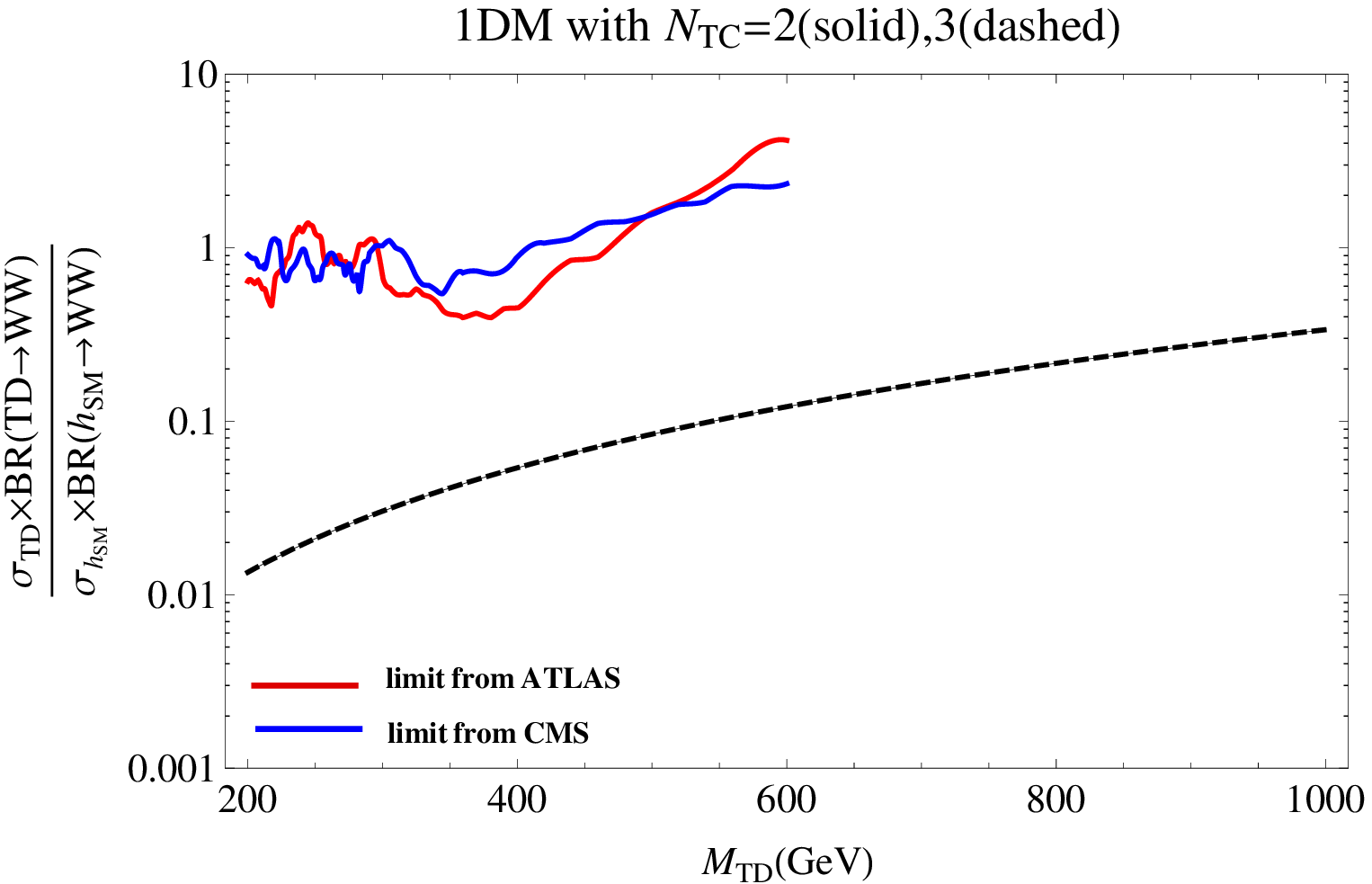}
 \includegraphics[width=8.5cm]{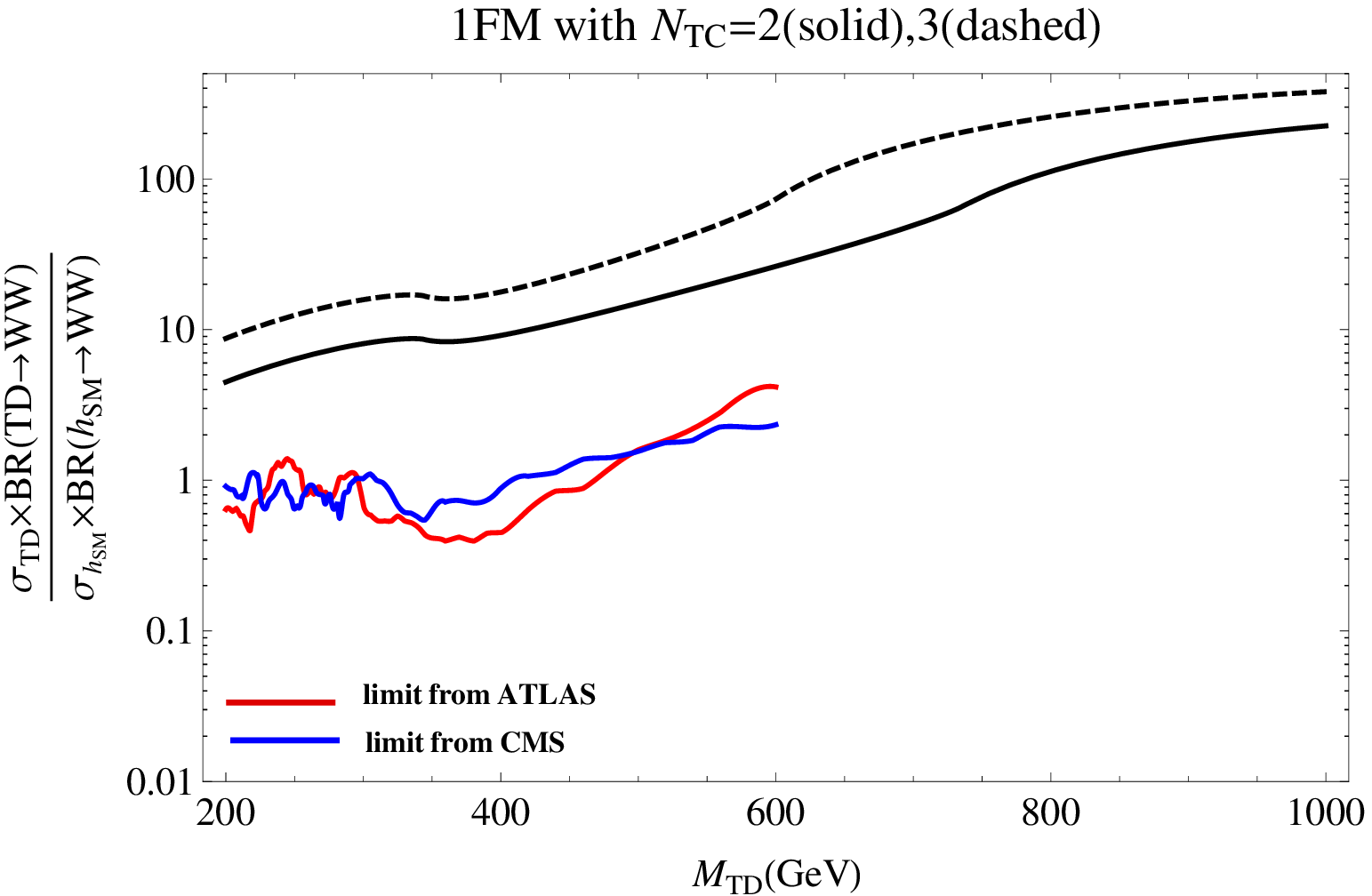}
\caption{ 
Left panel: The TD LHC production cross sections at $\sqrt{s}=7$ TeV times the $WW/ZZ$ branching ratio 
in the 1DMs with $N_{\rm TC}=2,3$ normalized to the corresponding quantity for the SM Higgs.  
Also shown is the comparison with the 95\% C.L. upper limits from the ATLAS~\cite{ATLAS} 
and CMS~\cite{CMS}. 
Right panel: 
The same as the left panel for the 1FMs.   
\label{LHCsignature1}
}
\end{center} 
 \end{figure}

We now check the consistency of the TD signatures 
 with the recent data at the LHC accumulated by the ATLAS and CMS detectors~\cite{ATLAS,CMS}. 
Varying the TD mass in the range $200 \, {\rm GeV} < M_{\rm TD} < 1000$ GeV,  
in Fig.~\ref{LHCsignature1} 
we plot the TD LHC production cross sections at $\sqrt{s}=7$ TeV  
times the $WW$ branching ratio ($R_{WW}$ in Eq.(\ref{R})) 
normalized to the corresponding quantity for the SM Higgs. 
The red and blue curves stand for the 95\% C.L. upper limits from the ATLAS and CMS experiments, respectively.  
  Looking at Fig.~\ref{LHCsignature1}, we see that in the case of 1DMs the TD signature is consistent with 
the current experimental data over all the mass range $200 \, {\rm GeV }< M_{\rm TD} < 600$ GeV 
thanks to the large suppression of GF cross section, %though
which on the other side of the coin would imply that  the TD may 
be invisible through this channel in contrast to the SM Higgs.   
 For the 1FMs, on the other hand, the consistency with the experimental data requires 
the TD mass to be $M_{\rm TD} \gtrsim$ 600 GeV, 
which conversely would imply that the TD can 
be discovered through somewhat large excesses at $600 \, {\rm GeV} \lesssim M_{\rm TD} <$ 1000 GeV 
in this channel in the near future.

\begin{figure}%[h]
\begin{center}
   \includegraphics[width=8.5cm]{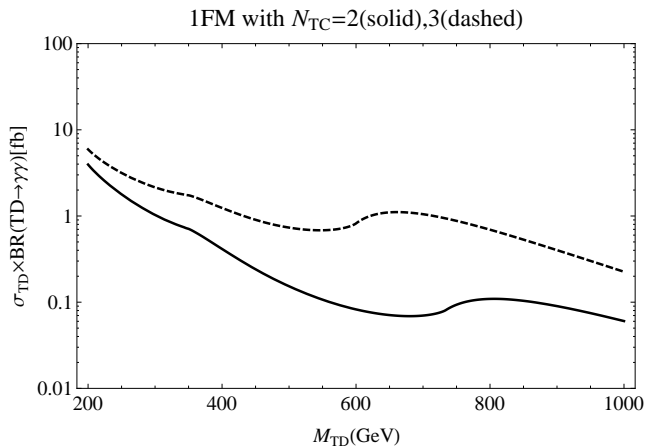}
\caption{ 
The TD  LHC production cross section  at $\sqrt{s}=7$ TeV times the $\gamma\gamma$ branching ratio in unit of fb 
for the 1FMs with $N_{\rm TC}=2,3$.  
\label{LHCsignature2}
}
\end{center} 
 \end{figure}

Since the $\gamma\gamma$ mode in the 1FMs get highly enhanced in contrast to the SM Higgs case 
as seen from Table~\ref{list:signals}, 
in Fig.~\ref{LHCsignature2} we finally plot   
the TD LHC production cross section at $\sqrt{s}=$7 TeV times the $\gamma\gamma$ branching ratio in the case of 1FMs 
over the TD mass range  $200\,  {\rm GeV} < M_{\rm TD} < 1000$ GeV.  
The figure tells us that the cross sections are large enough to be 
comparable with  the golden mode of SM Higgs signature 
$pp \to h_{\rm SM} \to ZZ \to l^+l^- l^+l^- \sim 1$ fb 
around the SM Higgs mass $\simeq $ 600 GeV: 
 At around $M_{\rm TD} \simeq $ 600 GeV, indeed, we have  
\begin{equation} 
\sigma_{\rm TD} \times BR({\rm TD} \to \gamma\gamma) \Bigg|_{\rm 1FM} 
\sim 0.10 \, (1.0) \, {\rm fb} 
\,,\qquad 
{\rm for} 
\qquad 
N_{\rm TC}=2(3)   
\,. \label{TDgammagamma:signal}
\end{equation}
This implies that the TD can be discovered through the $\gamma\gamma$ channel at the upcoming several months.

\section{Summary and discussion} 

In this paper we have studied the LHC signatures of TD 
arising as a composite pNGB of 
the spontaneous and explicit  breaking of the scale symmetry in WTC. 
The TD couplings to the SM particles were obtained, 
based on the  nonlinear realization of the electroweak symmetry as well as the scale symmetry (via  spurion method) which 
is broken not only spontaneously but also explicitly by the very origin of the spontaneous breaking, i.e., the dynamical generation of the techni-fermion mass (nonperturbative scale anomaly).
As a result, the TD couplings are similar to those of 
the SM Higgs except the overall scale set by the TD decay constant (and the anomalous dimension)
instead of the EW scale.

We took typical models of WTC such as 1DM and 1FM 
to make an explicit estimate of the branching ratios and production 
cross sections. 
To be more concrete, we further adopted the results from  
the recent LSD analysis combined with the PCDC relation  
to get the TD couplings as functions of the TD mass only.

We  calculated the TD decay widths and branching ratios 
to find that in the case of 1DMs the branching fraction becomes 
almost identical to that of the SM Higgs  
because of the similar form (though suppressed) of the 
TD coupling (See Eq.(\ref{Yukawa:ratio})). 
While in the 1FMs  the branching ratios of ${\rm TD} \to gg$ and $\gamma\gamma$ 
get highly enhanced due to presence of the extra contributions of techni-quarks (See Table~\ref{list:branching}). 
The same mechanism is effective also in the production cross sections:  
For 1DMs both the GF and VBF cross sections are suppressed compared 
to the SM Higgs case due to the suppression of TD couplings, 
while for 1FMs they get larger, where the amount of enhancement 
becomes more outstanding for the GF production (See Table~\ref{list}).

The TD LHC signatures were then discussed by explicitly calculating 
the production cross sections times the branching ratios 
in comparison with the corresponding quantities for the SM Higgs. 
 It turned out that the TD signatures look like quite different from the corresponding SM Higgs signatures: 
For 1DMs, all the TD signatures become 
one order of magnitude down compared with those of the SM Higgs, which would imply that if the Higgs-like object was found in the
region above 600 GeV, it would be naturally interpreted as TD, since the SM Higgs unlikely has such a large mass. 
On the other hand, for 1FMs all those signatures were shown to be enhanced by a factor of ${\cal O}(10)-{\cal O}(10^2)$ 
and would become outstanding signals of TD (See Table~\ref{list:signals}).

Varying the TD mass in the range $200 \, {\rm GeV} <M_{\rm TD} < 1000$ GeV, 
the cross section $pp \to {\rm TD} \to WW$ per the corresponding quantity 
for the SM Higgs has been compared 
with the recent upper limits from the ATLAS and CMS experiments.  
In the case of 1DMs the TD signature is consistent with 
the current experimental data over the mass range $200 \, {\rm GeV }< M_{\rm TD} < 600$ GeV 
thanks to the large suppression of GF cross section, 
though the TD may be invisible through this channel in contrast to the SM Higgs. 
For the 1FMs, on the other hand, the signals enhanced over the whole mass range 
 are severely constrained and the consistency with the experimental data requires 
the TD mass to be $M_{\rm TD} >600\,{\rm GeV}$, 
which, however, would also imply 
that the TD might be discovered through somewhat large excesses 
at $600 \, {\rm GeV} \lesssim M_{\rm TD} <$ 1000 GeV in this channel in the near future 
(See Fig.~\ref{LHCsignature1}).

Furthermore, the cross section $pp \to {\rm TD} \to \gamma\gamma$ for the 1FMs  
was predicted to be $\sim 0.10-1.0$ fb for the TD mass around 600 GeV 
(See Fig.~\ref{LHCsignature2} and Eq.(\ref{TDgammagamma:signal})).  
This cross section is large enough for the TD to be discovered in the upcoming several months 
and hence will also be a characteristic signature of TD.   
\\

Before closing this section, several comments are in order: 
\\

{\it  TD LHC signature from the $t\bar{t}$ mode} --- 
First of all, it is worth commenting on the TD LHC signature through the decay to $t\bar{t}$.   
 The ATLAS at $\sqrt{s}=7$ TeV with the integrated luminosity of $0.7$ fb$^{-1}$    
has already measured the $t\bar{t}$ production cross section $\sigma_{t\bar{t}}$ 
to report that $\sigma_{t\bar{t}} = 179.0 \pm 11.8$ pb~\cite{ATLAS:ttbar}, in agreement with 
the SM prediction. 
The SM Higgs contribution to this cross section 
is estimated to be about 0.1 pb at around the mass $\simeq 600$ GeV and hence 
is negligible in comparison with the dominant QCD contributions incorporated 
in the SM estimate of $\sigma_{t\bar{t}}$. 
As seen from Table~\ref{list:signals}, on the other hand, 
the cross section $\sigma(pp \to {\rm TD} \to t\bar{t})$ in the case of 1FMs gets enhanced 
by a factor of about 30 (70) for $N_{\rm TC}=2$(3) with $M_{\rm TD}=600$ GeV  
compared to the SM Higgs case, so we would have $\sigma(pp \to {\rm TD} \to t\bar{t}) \simeq 3(7)$ pb. 
Thus this signature is currently just as much as the size of the measurement uncertainties, 
but would be testable if more precise measurement of $\sigma_{t\bar{t}}$ becomes possible in the future. 
\\

{\it Perturbative unitarity} ---  
 For the 1FMs we might put an upper limit on $M_{\rm TD}$ coming from the perturbative unitarity 
bound through a formula, 
$
M_{\rm TD} \lesssim \Lambda_{\rm uni} = \sqrt{8\pi}F_\pi = \sqrt{8 \pi} \cdot \left( \frac{246\, {\rm GeV}}{\sqrt{N_{\rm D}}}\right)  
$~\cite{Lee:1977yc}.   
 By this formula we get $\Lambda_{\rm uni} \simeq 617$ GeV for the 1FMs, while it is absent for the 1DMs because 
 the bound is estimated to be above 1 TeV. 
Looking at Figs.~\ref{LHCsignature1} and \ref{LHCsignature2} with this unitarity bound taken into account,  
one might think that the 1FMs have completely been ruled out (up to the narrow window $600 \, {\rm GeV}<M_{\rm TD}<617\,{\rm GeV}$). 
Note however that the perturbative unitarity bound is thought of as just a reference, 
which may not make much sense, since WTC is itself of course unitary and 
the low-energy effective theory is still strongly coupled and not perturbative. 
\\

{\it Comparison with other dilaton phenomenology~\cite{Goldberger:2007zk}} ---  
Another approach on a dilaton signature at the LHC has been addressed in Ref.~\cite{Goldberger:2007zk}.   
Their dilaton is, however, completely different from our TD in the sense that theirs' is not due to
the scale symmetry of the WTC but the whole system including SM and the underlying EW theory 
(possibly including the WTC). 
Thus the characteristic feature is completely different from ours.
\\

{\it Non-running coupling case} ---  
In this paper we employed LSD calculation of the nonperturbative scale anomaly via vacuum energy, 
using the two-loop running coupling with CBZ-IRFP in Ref.~\cite{Hashimoto:2010nw}. 
We here discuss the comparison with the non-running limit. In this limit 
the  very origin of the dynamical mass $m_F$ comes from the cutoff $\Lambda$, 
the relation (hierarchy) between the two being given in the characteristic form of essential singularity 
(Miransky scaling)~\cite{Miransky:1984ef}: 
\begin{equation}
m_F \sim \Lambda\cdot \exp 
\left(
- \frac{\pi}{
\sqrt{
\frac{\alpha}{\alpha_c}-1
}}
\right)
\ll \Lambda \quad 
\left( {\alpha} \searrow {\alpha_c} \right)\, .
\label{Miransky}
\end{equation}
The  coupling $\alpha$ should depend on  $\Lambda/m_F$ in such a way that
$\alpha(\Lambda/m_F) \to \alpha_c$ in the limit $\Lambda/m_F  \to \infty$, resultant beta function being $\beta(\alpha) =\Lambda \partial \alpha/\partial \Lambda=-(2 \alpha_c /\pi)  (\alpha/\alpha_c-1)^{3/2}$. Hence the cutoff cannot be removed without requiring {\it nonperturbative running} of the coupling: 
The coupling (non-running in the perturbative sense) does actually run  slowly (walking) towards the critical coupling 
(regarded as the ultraviolet (UV) fixed point) at $\Lambda/m_F \rightarrow \infty$, 
reflecting the scale anomaly, i.e., explicit breaking of the scale symmetry (See Figs. 1(a) and (b) 
in the first of Ref.~\cite{Yamawaki:1985zg} for the beta function (a) and the associated anomalous dimension (b)).

Now in the WTC model based on the two-loop coupling with CBZ-IRFP, 
the role of the cutoff $\Lambda$ in the non-running case of the original WTC model~\cite{Yamawaki:1985zg,Bando:1986bg} 
is simply traded for  the intrinsic scale $\Lambda_{\rm TC} (\gg m_F)$ of the two-loop coupling, 
which breaks the scale symmetry already at perturbative level for the UV region ($p>\Lambda_{\rm TC}$) 
where the coupling runs as $1/\ln p$ as in the ordinary QCD.  
Although in the IR region ($p<\Lambda_{\rm TC}$) the coupling governed by CBZ-IRFP is almost non-running (scale invariant/conformal), 
the scale symmetry is broken both spontaneously and explicitly by yet another dynamics, namely the dynamical
generation of the techni-fermion mass $m_F (\ll \Lambda_{\rm TC})$,  and the coupling does run 
according to the nonperturbative renormalization a la Miransky in much the same way as the nonperturbative running of 
the non-running coupling case mentioned above.  It was  argued~\cite{Miransky:1996pd}  that  for the dynamical mass generation 
of essential singularity type scaling
Eq.(\ref{Miransky}),  characterized as ``conformal phase transition'',  
the associated nonperturbative scale anomaly (explicit breaking of the scale symmetry) is  
saturated by 
the pseudo dilaton (``massive dilaton'')  which are dictated by the PCDC.

Although the analysis of Ref.~\cite{Hashimoto:2010nw} was done for the case $\Lambda_{\rm ETC} \simeq\Lambda_{\rm TC}$,
we may take  a  choice $\Lambda_{\rm ETC} \ll \Lambda_{\rm TC}$, in which case  
the WTC would simply be reduced to the original WTC model of Ref.~\cite{Yamawaki:1985zg,Bando:1986bg} 
where the two-loop perturbative coupling becomes essentially scale invariant (non-running) 
 all the way up to $\Lambda=\Lambda_{\rm ETC}$~\footnote{
In this case WTC dynamics in isolation does not make sense in the asymptotically-free region 
$p>\Lambda_{\rm TC} (\gg \Lambda_{\rm ETC})$,  since then the theory will be  changed already 
at lower scale $\Lambda_{\rm ETC}$ into a different theory, ETC 
(or the preon model where SM particles and techni-fermions are the composites on the same footing~\cite{Yamawaki:1982tg}). }.  
In such a case the estimate of the scale anomaly through the vacuum energy was done long time ago~\cite{Miransky:1989qc} 
which differs from Eq.(\ref{scaleanomaly}) only by  the factor 0.81 in place of 0.76.  
Thus  our estimate in the text would be qualitatively the same as it stands:  
In fact, the estimated values of $m_F$ and $F_{\rm TD}$ in Eqs.(\ref{mF}) and (\ref{FTD}) 
will  be shifted upward only by about 5\% in the case of the non-running coupling. 
This happens because of about 5\% reduction and enhancement of the overall numerical coefficients 
in Eqs.(\ref{PS}) and (\ref{PCDC}), respectively: 0.41 $\to$ 0.39 in Eq.(\ref{PS}), while  
3.0  $\to$ 3.2 in Eq.(\ref{PCDC}). 
The Yukawa couplings will then get smaller than those listed in Table~\ref{list} by a factor of about 17\%, 
so the GF cross sections will be reduced by about 30\% at around $M_{\rm TD}=600$ GeV. 
The amount of TD LHC signatures will thus be reduced only by about 30\%  
compared to those predicted in Figs.~\ref{LHCsignature1} and \ref{LHCsignature2} and will not substantially change.

\section*{Acknowledgments}

We would like to thank T.~Abe and M.~Hashimoto for useful comments and discussions. 
This work was supported by 
the JSPS Grant-in-Aid for Scientific Research (S) \#22224003 and (C) \#23540300 (K.Y.).

\end{document}